\documentclass{article}
\usepackage{braket}
\usepackage{amsmath}

\begin{document}

\title{Binary Correlation Measurements}


\author{Toru Ohira\thanks{The author is also affiliated with Future Value Creation Research Center, Graduate School of Informatics, Nagoya University, and with Mathematical Science Team, RIKEN Center for Advanced Intelligence Project.}\\
Graduate School of Mathematics, Nagoya University, Nagoya, Japan} 

\maketitle

\begin{abstract}%
We search a simplest and minimal way to determine whether a given quantum system is entangled or separable. For this end, we propose binary correlation measurements in which restricted knowledge of only zero or non-zero correlations is available. We consider the concrete investigation on a pure state for two particles, each particle having two basis states (the $2\times 2$ system). We show that, even with this limited information from the binary correlation measurements, we can still reach the known minimum of three measurements for entanglement detection.
We next consider the comparable problem applied to the mixed density matrix. The mixed quantum case appears to require more detailed information, which we illustrate by studying the concrete example of the Werner density matrix.   
\end{abstract}

\section{Introduction}

Correlation measurements are a standard tool to investigate quantum systems. To determine whether the quantum system is separable or entangled, correlation measurements have been actively performed alongside of various theoretical investigations. (e.g.,\cite{bell,clauser,aspect,leggett,vedral,guhne03,altepeter,sakai}). For example, a recent work by Fujikawa et. al.\cite{fujikawa2} analyzed experimental results with the view of associating separability with zero--correlations, which is one of many active endeavors for obtaining conditions for separabilities of density matrices(e.g.,\cite{werner,peres,horodecki})

Here, we try to be as simplistic as possible and consider the approach of utilizing correlation functions that appear in traditional theories and measurements to detect entanglements. In particular, we base our analysis on achieving 
zero--correlations among observables measured on a composite quantum system. In other words, 
we assume that we need only measure the existence of correlations (either zero or non-zero), which we term
as the ``Binary Correlation Measurements''.
We do not require information on the exact values of non-zero correlation functions. It should be noted that this constraint simplifies the analysis of measurements.
This approach is different from other more elaborate mathematical methods for entanglement detections or separability conditions where more detailed information on correlation functions and/or the density matrix is necessary (\cite{horodecki2,guhne} for overviews).

The main question we proposed to ask with the above setup is the following:
what is the minimum number of the binary correlation measurements we need to perform to detect entanglements with certainty? To gain insights toward this question, we start with the simplest example of the two-particle (bipartite) $2\times 2$ system. 

First, we examine a pure quantum state. The known results for the minimal number of general correlation measurements are three to distinguish the entangled state from the separable one with certainty\cite{guhne03}.
We show that, even with the limited information obtained from the binary correlation measurements, three measurements involving four observables are still sufficient.

For the mixed case, however, the answer to the above question is not obtained. We illustrate the difficulty by considering the density matrix called the Werner density matrix that has a parameter whose value decides whether it is separable or entangled. We show that the correlation structure is the same regardless of whether there is entanglement or separability. Particularly, for this matrix, zero correlations appear for the entangled situation just in the same way as the separable situation. This implies that non-zero correlations do not certify the existence of entanglement nor does zero correlation entail separability.

We end the paper with discussions on extensions of this line of approach to higher dimensions as
 well as general questions on concepts of correlations and entanglements in capturing quantum nature through a brief mention of the quantum pigeon hole effect.

\section{Main Question}

We consider a typical situation of correlation measurements on pure or mixed quantum systems. Let us start with the simplest case of
a bipartite system consisting of two quantum particles $A$ and $B$. We can measure observable operators $\mathcal{X}$ and $\mathcal{Y}$ defined as
$\mathcal{X} = {\mathcal{Q}_A}\otimes{\bf{1}}_B$ and $\mathcal{Y} = {\bf{1}}_A\otimes\mathcal{R}_B$. They can be measured
independently meaning measurement only on particle $A$, or only on particle $B$. Also, we want to measure
$\mathcal{X}\mathcal{Y}$, both measurements necessarily taking place at the same time. Through repeating these measurements for statistics, we obtain three expectation values, $\bra{\psi}\mathcal{X}\ket{\psi}, \bra{\psi}\mathcal{Y}\ket{\psi}$, and $\bra{\psi}\mathcal{X}\mathcal{Y}\ket{\psi}$, from which we can compute the correlation (covariance)
as
\begin{equation}
c(\mathcal{X}, \mathcal{Y}) = \bra{\psi}\mathcal{X}\mathcal{Y}\ket{\psi} - \bra{\psi}\mathcal{X}\ket{\psi}\bra{\psi}\mathcal{Y}\ket{\psi}.
\end{equation}

As is well known, if the system is described by a density matrix $\rho_{AB}$, we can generalize the above using its trace and its partial traces $\rho_A$ and $\rho_B$ to describe the quantum state and compute the correlation function in the following manner.
\begin{equation}
c(\mathcal{X}, \mathcal{Y}) = Tr(\rho_{AB}\mathcal{X}\mathcal{Y}) - Tr(\rho_{A}\mathcal{X})Tr(\rho_{B}\mathcal{Y}).
\end{equation}

We also mention that a density matrix describing a pure quantum system is separable if it can be written as
\begin{equation}
\rho_{AB}=\rho_A\otimes\rho_B,
\end{equation}
otherwise it is entangled. For a density matrix describing a mixed state, the separability is defined as
\begin{equation}
\rho_{AB}=\sum_i p_i\rho^i_{A}\otimes\rho^i_{B}
\label{mseparable}
\end{equation}
with non-negative real $\{ p_i \}$ such that $\sum_i p_i = 1$. If such decomposition is
not possible, it is entangled. 

\vspace{1em}
With the set up above our main question can be phrased as 
\vspace{1em}

\noindent
{\bf{Main Question}}

What is the minimum number of binary correlation measurements we need to perform on a given quantum state to determine with certainty whether it is entangled or separable?
\vspace{1em}

We assume that we can design experiments to measure observables $\mathcal{X}$ and $\mathcal{Y}$ to obtain the above mentioned-expectation values. The question is to find the minimum number of measurements of correlation functions so that we can be certain to determine whether the given system is in a separable or an entangled state. 
We also assume that the correlation measurement results are binary: either zero or non-zero. That is to say, we do not need to know the values of any non--zero correlation function. 
We will see that this leads to a relatively simple measurement analysis. 
It should be noted that, with this restriction, our approach is different from much investigated separability criteria and entanglement detections, where more information from quantum systems is required\cite{horodecki2,guhne}.
In the following, we illustrate how much 
we can gain insight into quantum systems even with these restricted correlation measurements.

\section{Analysis with the $2\times 2$ system}

To gain insight, we proceed to consider the simplest bipartite system of two 2-state particles ($2\times 2$ system), such as two spin 1/2 particles or two qubits systems(e.g.,\cite{wootters,kummer,abouraddy,jchen}).

Our analysis will employ the density matrix, whether or not the system is in
a pure or mixed state.
It is known that any density matrix for $2\times2$ systems can be written using the Pauli matrices as follows.

\begin{equation}
\rho_{AB}={1\over 4}({\bf{1}_A}\otimes {\bf{1}_B}+{\vec{a}}\cdot{\vec{\mathcal{\sigma}}} \otimes {\bf{1}_B} + {\bf{1}_A}\otimes {\vec{b}}\cdot{\vec{\mathcal{\sigma}}} +\sum_{ij} F_{ij}{\mathcal{\sigma}}_i\otimes {\mathcal{\sigma}}_j)
\label{density}
\end{equation}
where ${\bf{1}}$ is the $2\times2$ identity matrix, ${\vec{a}},{\vec{b}}$ are vectors consists of 3 real numbers ($\cdot$ is the inner product),
and $F_{ij}$ are real number elements of a $3\times3$ matrix ${\mathcal{F}}$,
and ${\vec{\mathcal{\sigma}}} = ({\mathcal{\sigma}_x}, {\mathcal{\sigma}_y},{\mathcal{\sigma}_z})$ is the vector with the Pauli matrices.
\begin{equation}
{\mathcal{\sigma}_x}=
\begin{bmatrix}
0 & 1 \\
1 & 0
\end{bmatrix}
, \quad
{\mathcal{\sigma}_y}=
\begin{bmatrix}
0 & -i \\
i & 0
\end{bmatrix}, \quad
{\mathcal{\sigma}_z}=
\begin{bmatrix}
1 & 0 \\
0 & -1
\end{bmatrix}.\nonumber
\end{equation}

Now, the two quantum observable operators ${\mathcal{Q}_A}, {\mathcal{R}_B}$ can be also expressed using the Pauli matrices up to a scale factor as
\begin{equation}
{\mathcal{Q}_A} = {1\over 2}({\bf{1}_A} + {\vec{x}}\cdot{\vec{\mathcal{\sigma}}}), \quad {\mathcal{R}_B} = {1\over 2}({\bf{1}_B} + {\vec{y}}\cdot{\vec{\mathcal{\sigma}}})
\label{obs}
\end{equation}
where ${\vec{x}},{\vec{y}}$ are three dimensional real vectors.

Then, for operators $\mathcal{X} = {\mathcal{Q}_A}\otimes{\bf{1}}_B$ and $\mathcal{Y} = {\bf{1}}_A\otimes\mathcal{R}_B$, we can calculate the expectation values as follows;
\begin{equation}
\langle \mathcal{X}\mathcal{Y} \rangle = {\mathrm{Tr_{AB}}}[\rho_{AB}\mathcal{X}\mathcal{Y}]={1\over 4}(1+
{\vec{a}}\cdot{\vec{x}} + {\vec{b}}\cdot{\vec{y}} + {\vec{x}}\cdot{\mathcal{F}}\cdot{\vec{y}}),\nonumber
\end{equation}
and
\begin{equation}
\langle \mathcal{X} \rangle ={1\over 2}(1 + {\vec{a}}\cdot{\vec{x}}),\quad 
\langle \mathcal{Y} \rangle ={1\over 2}(1 + {\vec{b}}\cdot{\vec{y}}). \nonumber
\end{equation}
This leads to
\begin{equation}
\langle \mathcal{X}\mathcal{Y} \rangle - \langle \mathcal{X} \rangle\langle \mathcal{Y} \rangle = {1\over 4}({\vec{x}}\cdot{\mathcal{F}}\cdot{\vec{y}} - ({\vec{a}}\cdot{\vec{x}})({\vec{b}}\cdot{\vec{y}})) = {1\over 4}{\vec{x}}\cdot({\mathcal{F}} - {\vec{a}}\cdot{{\vec{b}^{\hspace{0.05cm} \mathsf{T}}}})\cdot{\vec{y}},
\end{equation}
where ${\vec{a}}\cdot{{\vec{b}^{\hspace{0.05cm} \mathsf{T}}}}$ is the outer product of ${\vec{a}}, {{\vec{b}}}$. 
By defining a real $3\times3$ matrix $\mathcal{C}$ as
\begin{equation}
\mathcal{C} = {\mathcal{F}} - {\vec{a}}\cdot{{\vec{b}^{\hspace{0.05cm} \mathsf{T}}}},
\label{corr}
\end{equation}
\begin{equation}
c(\mathcal{X}, \mathcal{Y}) = {1\over 4}{\vec{x}}\cdot \mathcal{C} \cdot{\vec{y}}.
\label{corrv}
\end{equation}
Thus, the nature of the correlation function depends on that of the matrix $\mathcal{C}$.
Particularly, we note the following\cite{cheng}, which generalizes our previous analogous results\cite{ohira1,ohira2} for any
pure bipartite $2\times2$ state vectors. 
\vspace{1em}

\noindent
{\bf{Theorem 1}}:
For any bipartite $2\times2$ system  (regardless of separable or entangled, or, pure or mixed), there always exits a pair of observables
$\mathcal{X}, \mathcal{Y}$ that gives a zero--correlation, $c(\mathcal{X}, \mathcal{Y}) = 0$.
\vspace{1em}

\noindent
{\bf{Proof}}: From the setup above, zero--correlation is achieved when ${\vec{x}}\cdot \mathcal{C} \cdot{\vec{y}} = 0$. This is simply an orthogonality relation of two real vectors ${\vec{x}}$ and $\mathcal{C} \cdot{\vec{y}}$ in three-dimensional space. Such a pair of ${\vec{x}}, {\vec{y}}$ can always be found for any real $3\times3$ matrix $\mathcal{C}$. (Q.E.D)
\vspace{1em}

This proof shows that a zero--correlation observable pair can not only be found for any $\rho_{AB}$, but also there are uncountably many such pairs. 
It also turns out that, for pure states, the special nature of the matrix $\mathcal{C}$ allows us to answer our main question of distinguishing 
entangled states.

\subsection{Pure States}

For the case of pure states, the density matrix has special properties, which in turn constrain
the nature of the matrix $\mathcal{C}$.
\vspace{1em}

\noindent
{\bf{Theorem 2}}: For a pure state, the rank of the $3\times3$ matrix $\mathcal{C}$ is given as follows

(a) For a separable state $rank(\mathcal{C}) = 0$, that is $\mathcal{C}$ is the zero matrix.

(b) For an entangled state $rank(\mathcal{C}) = 3$, that is $\mathcal{C}$ is a regular matrix.
\vspace{1em}

\noindent
{\bf{Proof}}: First, we list some known properties\cite{jchen} when the density matrix in (\ref{density}) describes a pure state.

(i) ${\mathcal{F}}\cdot \vec{b} = \vec{a}, \quad {\mathcal{F}}^{\hspace{0.05cm} \mathsf{T}}\cdot \vec{a} = \vec{b}$.

(ii) $0 \leq \begin{Vmatrix} \vec{a} \end{Vmatrix} = \begin{Vmatrix} \vec{b} \end{Vmatrix} \leq 1$ ($\vec{a}$ and $\vec{b}$ has the same length).

(iii) $Det({\mathcal{F}}) = \begin{Vmatrix} \vec{a} \end{Vmatrix}^2 - 1 = \begin{Vmatrix} \vec{b} \end{Vmatrix}^2 -1$.

(iv)The density matrix describes a pure and separable state if and only if $\begin{Vmatrix} \vec{a} \end{Vmatrix} = \begin{Vmatrix}\vec{b} \end{Vmatrix} = 1$.
\vspace{1em}

\noindent
(a) We note that any density matrix describing a pure separable state can be written as
\begin{equation}
\rho^{sep}_{AB} = {1\over 2}({\bf{1}_A} + {\vec{a}}\cdot{\vec{\mathcal{\sigma}}})\otimes {1\over 2}({\bf{1}_B} + {\vec{b}}\cdot{\vec{\mathcal{\sigma}}})
\label{sep}
\end{equation}
By expanding this, we can immediately see with (\ref{density}) that 
\begin{equation}
{\mathcal{F}} = {\vec{a}}\cdot{{\vec{b}^{\hspace{0.05cm} \mathsf{T}}}}.
\end{equation}
Hence, by the definition of the correlation matrix (\ref{corr}),
\begin{equation}
\mathcal{C} = {\mathcal{F}} - {\vec{a}}\cdot{{\vec{b}^{\hspace{0.05cm} \mathsf{T}}}} = \mathcal{O}.
\end{equation}
\vspace{1em}

\noindent
(b) We consider two possible cases.
\vspace{1em}

\noindent
For ${\vec{a}} = {\vec{b}} = {\vec{0}}$: By the definition (\ref{corr}) of the correlation matrix $\mathcal{C} = \mathcal{F}$, and therefore by the property (iii), we have 
\begin{equation}
Det({\mathcal{C}}) = Det({\mathcal{F}}) =  - 1
\end{equation}
This shows that the correlation matrix is regular and thus has $rank(\mathcal{C}) = 3$.
\vspace{1em}

\noindent
For ${\vec{a}} \neq {\vec{0}}$: By the property (ii), we also have ${\vec{b}} \neq {\vec{0}}$. 
Without loss of generality, we can choose a coordinate system such that
\begin{equation}
{\vec{b}} = \begin{Vmatrix}\vec{b} \end{Vmatrix}\begin{bmatrix}
1 \\
0\\
0
\end{bmatrix}
\end{equation}
Then, by the property (i),
\begin{equation}
\vec{a} = {\mathcal{F}}\cdot \vec{b}
= \begin{Vmatrix}\vec{b} \end{Vmatrix} \begin{bmatrix}
F_{11} \\
F_{21}\\
F_{31}
\end{bmatrix}.
\end{equation}

We can now derive the following by (\ref{corr}) and the property of the determinant,
\begin{equation}
Det(\mathcal{C}) = Det(
\begin{bmatrix}
(1-\begin{Vmatrix}\vec{b} \end{Vmatrix}^2)F_{11},& F_{12},& F_{13} \\
(1-\begin{Vmatrix}\vec{b} \end{Vmatrix}^2)F_{21},& F_{22},& F_{23} \\
(1-\begin{Vmatrix}\vec{b} \end{Vmatrix}^2)F_{31},& F_{32},& F_{33} 
\end{bmatrix})
= (1-\begin{Vmatrix}\vec{b} \end{Vmatrix}^2)Det({\mathcal{F}}) = - (\begin{Vmatrix} \vec{b} \end{Vmatrix}^2 -1)^2.
\end{equation}
The right-hand side of this equation is non-zero by the property (iv) for the entangled case, which, in turn, shows that the correlation matrix is regular and thus has $rank(\mathcal{C}) = 3$. (It also validates and includes the case of ${\vec{a}} = {\vec{b}} = {\vec{0}}$.) (Q.E.D)
\vspace{1em}

This theorem provides an answer to our main question to find the minimum number of the binary correlation measurements for entanglement detection.
\vspace{1em}

\noindent
{\bf{Theorem 3}}: For pure bipartite $2\times2$ quantum systems, the minimum number of the binary correlation measurements for entanglement detection with certainty is three.
\vspace{1em}

\noindent
{\bf{Proof}}: From Eq. (\ref{corrv}), and Theorem 2(a), the correlation function $c(\mathcal{X}, \mathcal{Y})$ is $0$ for the separable case for any pair of 
$\mathcal{X}, \mathcal{Y}$. Now, for the entangled state, let us choose $\mathcal{Y}$ to have any value other than the identity (no measurement). This is equivalent to fixing a real 
three-dimensional vector 
$\vec{y}(\neq 0)$. Then, the correlation function is $0$ if and only if the vector $\vec{x}$ defining $\mathcal{X}$ lies in the plane perpendicular to
${\vec{y'}}\equiv \mathcal{C} \cdot{\vec{y}}$ that is a non-zero vector by Theorem 2(b) for the entangled case. Thus, if we prepare thee observables
$\mathcal{X}_a, \mathcal{X}_b, \mathcal{X}_c$ defined by linearly independent\footnote{they do not have to be orthogonal.} $\vec{x_a}, \vec{x_b}, \vec{x_c}$, at least one of them 
gives non-zero correlation with respect to the fixed $\mathcal{Y}$. (The perpendicular plane can only accommodate up to two linearly independent real three-dimensional vectors.) 
This will identify the entangled states, distinguishing them from separable ones.  (Q.E.D)
\vspace{1em}

\noindent
{\bf{Examples}}:
To illustrate the above theorem, we discuss here two examples of pure entangled states.
\vspace{1em}

\noindent
(i) The first example is the singlet state, which is also one of the Bell states.  
\begin{equation}
\ket{\Psi^{-}}= {1\over\sqrt{2}}(\ket{a_1}\otimes\ket{b_2} - \ket{a_2}\otimes\ket{b_1})
\equiv {1\over\sqrt{2}}(\begin{bmatrix}
1 \\
0
\end{bmatrix}_A \otimes
\begin{bmatrix}
0 \\
1
\end{bmatrix}_B 
- 
\begin{bmatrix}
0 \\
1
\end{bmatrix}_A
\otimes
\begin{bmatrix}
1 \\
0
\end{bmatrix}_B). 
\label{bell2}
\end{equation}

The corresponding density matrix is given as follows in the notation
above:
\begin{equation}
\rho_{AB} ={1\over 4}({\bf{1}_A}\otimes {\bf{1}_B} + \sum_{i} (-1){\mathcal{\sigma}}_i\otimes {\mathcal{\sigma}}_i)
=  {1\over 2}\begin{bmatrix}
0,& 0,& 0,& 0 \\
0,& 1,& -1,& 0 \\
0,& -1,& 1,& 0 \\
0,& 0,& 0,& 0 
\end{bmatrix}
\label{density-singlet}
\end{equation}
Thus, in this case, the correlation matrix is particularly simple as
\begin{equation}
\mathcal{C} = \mathcal{F}
=  \begin{bmatrix}
-1,& 0,& 0 \\
0,& -1, & 0 \\
0,&  0, & -1 
\end{bmatrix}
\label{cmatrix-singlet}
\end{equation}
Thus, the correlation function (\ref{corrv}) is  $-{1\over 4}\vec{x}\cdot\vec{y}$, which is zero for any pair of observable $\mathcal{X}, \mathcal{Y}$ defined by a pair of orthogonal vectors $\vec{x}, \vec{y}$. Therefore, for a given $\mathcal{Y}$, it suffices
to prepare and measure three observables operators $\mathcal{X}_a, \mathcal{X}_b, \mathcal{X}_c$ defined by three linearly independent vectors $\{\vec{x_a}, \vec{x_b}, \vec{x_c}\}$: at least one of them will give a non-zero value, showing that this is an entangled
state.
\vspace{1em}

\noindent
(ii) The second example is from \cite{jchen}. The entangled state vector is given as 
\begin{eqnarray}
\ket{\Phi}&=&{1\over\sqrt{3}}(\ket{a_1}\otimes\ket{b_1} +\ket{a_1}\otimes\ket{b_2}+ \ket{a_2}\otimes\ket{b_2})\nonumber\\
&\equiv& {1\over\sqrt{3}}(\begin{bmatrix}
1 \\
0
\end{bmatrix}_A \otimes
\begin{bmatrix}
1 \\
0
\end{bmatrix}_B 
+ 
\begin{bmatrix}
1 \\
0
\end{bmatrix}_A
\otimes
\begin{bmatrix}
0 \\
1
\end{bmatrix}_B
+ 
\begin{bmatrix}
0 \\
1
\end{bmatrix}_A
\otimes
\begin{bmatrix}
0 \\
1
\end{bmatrix}_B). 
\label{chen}
\end{eqnarray}
The corresponding density matrix is given as
\begin{equation}
\rho_{AB}=\ket{\Phi}\bra{\Phi} = {1\over 9}\begin{bmatrix}
1,& 2,& 0,& 2 \\
2,& 4,& 0,& 4 \\
0,& 0,& 0,& 0 \\
2,& 4,& 0,& 4 
\end{bmatrix}
\label{density-then}
\end{equation}
This density matrix can be written in the form of (\ref{density}) with
\begin{equation}
{\vec{a}}={1\over 3}
\begin{bmatrix}
2 \\
0 \\
1
\end{bmatrix}
, \quad
{\vec{b}}={1\over 3}
\begin{bmatrix}
2 \\
0 \\
-1
\end{bmatrix}, \quad
{\mathcal{F}} =
\begin{bmatrix}
2 & 0 & -2\\
0 & -2 & 0 \\
2 & 0 & 1
\end{bmatrix}.\nonumber
\end{equation}
This yields the correlation matrix as
\begin{equation}
\mathcal{C} = {\mathcal{F}} - {\vec{a}}\cdot{{\vec{b}^{\hspace{0.05cm} \mathsf{T}}}}
= {2\over 9}
\begin{bmatrix}
1 & 0 & -2\\
0 & -3 & 0 \\
2 & 0 & 2
\end{bmatrix}.
\end{equation}
If we set 
\begin{equation}
{\vec{y}}=
\begin{bmatrix}
1 \\
0 \\
0
\end{bmatrix},\quad
{\vec{y'}} = \mathcal{C} \cdot{\vec{y}}={2\over 9}
\begin{bmatrix}
1 \\
0 \\
2
\end{bmatrix}.
\end{equation}
Thus, the correlation function is zero for any operator $\mathcal{X}$ defined by a vector ${\vec{x}}$ that resides on
the plane spanned by
\begin{equation}
\{
\begin{bmatrix}
2 \\
0 \\
-1
\end{bmatrix},
\begin{bmatrix}
0 \\
1 \\
0
\end{bmatrix}
\}
\end{equation}
Thus, again, as we prepare three observable operators defined by three linearly independent vectors $\{\vec{x_a}, \vec{x_b}, \vec{x_c}\}$, we can be sure that at least one of them
will yield a non-zero correlation.

\subsection{Mixed Density Matrix}

As has been known, if the system is described by a mixed density matrix, the situation is more complex. It is also the case for our approach here. We illustrate this by considering a concrete example of the Werner density matrix\cite{werner}.

The Werner density matrix for bipartite $2\times2$ quantum systems is given as follows, 
\begin{equation}
\rho^{W}_{AB}={{1-\xi}\over 4}{\bf{1}_A}\otimes {\bf{1}_B} + \xi\ket{\Psi^{-}}\bra{\Psi^{-}}
=  {1\over 4}\begin{bmatrix}
1-\xi,& 0,& 0,& 0 \\
0,& 1+\xi,& -2\xi,& 0 \\
0,& -2\xi,& 1+\xi,& 0 \\
0,& 0,& 0,& 1-\xi 
\end{bmatrix},
\end{equation}
where $\ket{\Psi^{-}}$ is the singlet state (\ref{bell2}), to which it reduces with $\xi=1$. Thus, it can be viewed
as a generalization of the singlet state. It can be rewritten as follows
\begin{equation}
\rho^{W}_{AB}={{1}\over 4}({\bf{1}_A}\otimes {\bf{1}_B} +(-\xi) \sum_{i}{\mathcal{\sigma}}_i\otimes {\mathcal{\sigma}}_i),
\label{mixdensity}
\end{equation}
which is of the form (\ref{density}), with $\vec{a}=\vec{b}=\vec{0}$, $F_{ij}= -\delta_{ij}\xi$. 
This density matrix has been studied extensively (e.g., \cite{peres,hiroshima,azuma} and known to be separable, i.e., it can be rewritten in the form of (\ref{mseparable}), for $0 \leq \xi \leq {1\over 3}$, and entangled for ${1\over 3} < \xi \leq 1$. 

Now, if we apply our procedure, the matrix $\mathcal{C}$ is a simple
diagonal matrix, which again reduces to that of the singlet state with $\xi=1$.
\begin{equation}
\mathcal{C} = \mathcal{F}
=  (-\xi)\begin{bmatrix}
1,& 0,& 0 \\
0,& 1,& 0 \\
0,& 0,& 1 
\end{bmatrix}
\label{cmatrix-werner}
\end{equation}
Hence,the correlation function is also simply given as
\begin{equation}
c(\mathcal{X}, \mathcal{Y}) = -{\xi\over 4}{\vec{x}}\cdot {\vec{y}}.
\end{equation}
This shows that the $c(\mathcal{X}, \mathcal{Y})=0$ if and only if
non-zero real three-dimensional vectors are orthogonal regardless
of the value of $\xi$. Thus, for the Werner density matrix, the zero-correlation condition is the same both for separable and entangled situations.
Zero correlations do not entail separability, nor do non-zero correlations mean entanglements. 
This is in contrast to the pure state case where zero-correlations can distinguish between 
the separable and entangled states. In general,
for the mixed density matrix, it appears that we need to analyze further the values of the correlation function with various measurements and apply other criteria to detect entanglements.

\section{Discussion}

\noindent
(i) Generalization

For pure $2 \times 2$ states, we expect the binary correlation measurements proposed here to give the ``minimal bound'' for entanglement detection.  For $n \times n$ states, naive extended conjecture is the following.
\vspace{1em}

\noindent
{\bf{Conjecture}}: For a bipartite $n \times n$ pure system, the following holds.

(1) The analogously defined correlation matrix $C$ has a rank of either $0$ for separable states, or $n^2 -1$ for entangled states.

(2) The minimum number of the correlation function measurements for entanglement detection with certainty is $n^2 -1$.
\vspace{1em}

It turns out that this conjecture is not quite correct. However, it paved the way to develop the first ever algorithm for computing the Schmidt rank of any unknown pure quantum state using only zero-correlation tests\cite{tanasescu}. Also for the $n \times n$  system,  a known minimum bound of $n+1$ for entanglement detections using general correlation measurements has been obtained\cite{guhne03}.  It is yet to be explored if our restricted binary correlation measurements can reach or approximate this bound.

For a multi-party system involving more than two, we cannot anticipate any analogous theorem. It is not clear even whether there exist entangled states that never admit zero correlations.
\vspace{1em}

\noindent
(ii) Limitations

As we have noted, this approach of correlation measurements particularly based on zero-correlation has not been found to be a useful way to distinguish separability from entanglement for mixed states in general. Even the existence of uncountably many zero-correlations does not certify separability, nor non-zero correlations entail entanglement. It appears that we need to gain more detailed information on the values of the correlation functions, or each element of the density matrix in question, from which more elaborate mathematical tools can be applied for entanglement determination. 
\vspace{1em}

\noindent
(iii) Prospects

Thus, the concept of correlations as we have set it up above is not enough to capture the nature of quantum entanglements in general. Further, we would like to point out that the concept of entanglement itself may not be enough to describe quantum correlations in general. This is indicated by the recently proposed ``quantum pigeonhole effect'' by Aharonov et. al. \cite{aharonov}. 

Classically the pigeonhole principle states that if we have more pigeons placed in fewer number of holes, at least
one hole must have multiple pigeons together. The analogous system is considered in quantum mechanics with three two-state quantum particles (pigeons), each in a superposition of two (hole) states (a quantum $3\times2$ system). Computing correlations with cleverly chosen different pre- and post-selected product (non-entangled) states, surprisingly, show that no pair of particles can be in the same quantum (hole) state. This means that the pigeonhole principle in some cases breaks down in quantum mechanics. It also shows there are new aspects of quantum entanglement which are not apparent in product states. The experimental work with three single photons transmitted through two polarization channels indicates that this quantum pigeonhole effect is real \cite{chen}.
Concepts, which may be a generalization of entanglement, to capture this type of quantum correlations are yet to be explored.  


\section*{Acknowledgment}

The author would like to thank Philip M. Pearle, Professor Emeritus of Hamilton College, for his comments and encouragements.
Also, comments by Dr. Shuming Cheng on Theorem 1, and by Profs. Chigaku Itoi and Shinichi Deguchi of Nihon University are acknowledged as very constructive for this work.
This work was supported by funding from Ohagi Hospital, Hashimoto, Wakayama, Japan, and by Grant-in-Aid for Scientific Research from Japan Society for the Promotion of Science No.19H01201.




\begin{thebibliography}{9}

\bibitem{bell}
 J. S. Bell, 
Physics, {\bf 1}, 195 (1964).


\bibitem{clauser}
J.F. Clauser, M.A. Horne, A. Shimony, and R.A. Holt, 
Phys. Rev. Lett., {\bf 23}, 880 (1969).

\bibitem{aspect}
A. Aspect, P. Grangier, and G. Roger,
Phys. Rev. Lett., {\bf 47}, 460 (1981).


\bibitem{leggett}
A. J. Leggett, and Anupam Garg,
Phys. Rev. Lett., {\bf 54}, 857 (1985).


\bibitem{vedral}
V. Vedral, M. B. Plenio, M. A. Pippin, and P. L. Knight,
Phys. Rev. Lett., {\bf 78}, 2257 (1997).

\bibitem{guhne03}
O. G{\"u}hne, P. Hyllus , D. Brus{$\beta$} , A. Ekert , M. Lewenstein , C. Macchiavello, and A. Sanpera,
J. of Modern Optics, {\bf 50}, 1079 (2003). 

\bibitem{altepeter}
J. B. Altepeter, E. R. Jeffrey, P. G. Kwiat, S. Tanizilli, N. Gisin, and A. Acin,
Phys. Rev. Lett., {\bf 95}, 033601 (2005).

\bibitem{sakai}
H. Sakai, T. Saito, T. Ikeda, K. Itoh, T. Kawahata, H. Kuboki, Y. Maeda, N. Matsui, C.Rangacharyulu, M. Sasano, Y. Satou, K. Sekiguchi, K. Suda, A. Tamio, T. Uesaka, and K. Yako,
Phys. Rev. Lett., {\bf 97}, 150405 (2006).

\bibitem{fujikawa2}
K. Fujikawa, C. H. Oh, K. Umetsu and S. Yu,
Annals of Physics, {\bf 368}, 248 (2016).




\bibitem{werner}
R. F. Werner,
Phys. Rev. A, {\bf 40}, 4277 (1989).


\bibitem{peres}
A. Peres,
Phys. Rev. Lett., {\bf 77}, 1413 (1996).

\bibitem{horodecki}
M. Horodecki, P. Horodecki, and R. Horodecki,
Rev. Lett. A, {\bf 223}, 1 (1996).



\bibitem{horodecki2}
R. Horodecki, P. Horodecki, M. Horodecki, and K. Horodecki,
Rev. Mod. Phys., {\bf 81}, 865 (2009).

\bibitem{guhne}
O. G{\"u}hne, G. T{\'u}th,
Physics Reports, {\bf 474}, 1 (2009).



\bibitem{wootters}
W. K. Wootters,
Phys. Rev. Lett., {\bf 80}, 2245 (1998).

\bibitem{kummer}
H. J. Kummer, 
Int. J. of Theor. Phys., {\bf 40}, 1071 (2001).

\bibitem{abouraddy}
A. F. Abouraddy, R. E. A. Saleh, A. V. Sergienko, and M. C. Fesch, 
Phys. Rev. A, {\bf 64}, 050101(R) (2001).

\bibitem{jchen}
J. L Chen, L. Fu, A. A. Ungar, and X, G Zhao,
Phys. Rev. A, {\bf 65}, 044303 (2002).


\bibitem{cheng}
S. Cheng (Private communication).

\bibitem{ohira1}
T. Ohira, 
Prog. Theor. Phys., {\bf 2018-8}, 083A02 (2018).

\bibitem{ohira2}
T. Ohira, 
Prog. Theor. Phys., {\bf 2020-1}, 013A01 (2020).


\bibitem{hiroshima}
T. Hiroshima and S. Ishizaka,
Phys. Rev. A, {\bf 62}, 044302 (2000).

\bibitem{azuma}
H. Azuma and M. Ban,
Phys. Rev. A, {\bf 73}, 032315 (2006).

\bibitem{tanasescu}
A. Tanasescu, A. Balan, P. G. Popescu,
European Physical Journal Plus, {\bf 136}, 476 (2021).


\bibitem{aharonov}
Y. Aharonov, F. Colombo, S. Popescu, I. Sabadini, D. C. Struppa and J. Tollaksen,
Proc. Natl. Acad. Sci. USA, {\bf 113}, 532 (2016).

\bibitem{chen}
M. Chen, C. Liu, Y. Luo, H. Huang, B. Wang, X. Wang, L. Li, N. Liu, and C. Lu and J. Pan,
Proc. Natl. Acad. Sci. USA, {\bf 116}, 1549 (2019).





\end{thebibliography}
%

\end{document}